# Opinion Formation Threshold Estimates from Different Combinations of Social Media Data-Types

Derrik E. Asher
US ARL
derrik.e.asher.civ@mail.mil

Justine Caylor
US ARL
justine.p.caylor.ctr@mail.mil

Casey Doyle
RPI
doylec3@rpi.edu

Alexis R. Neigel
US ARL
alexis.r.neigel.ctr@mail.mil

Gyorgy Korniss
RPI
korniss@rpi.edu

Bolek K. Szymanski
RPI
boleslaw.szymanski@gmail.com

US Army Research Laboratory (US ARL), Aberdeen Proving Ground, MD
Rensselaer Polytechnic Institute (RPI), Troy, NY

## Abstract

*Passive consumption of a quantifiable amount of social media information related to a topic can cause individuals to form opinions. If a substantial amount of these individuals are motivated to take action from their recently established opinions, a movement or public opinion shift can be induced independent of the information's veracity. Given that social media is ubiquitous in modern society, it is imperative that we understand the threshold at which social media data results in opinion formation. The present study estimates population opinion formation thresholds by querying 2222 participants about the number of various social media data-types (i.e., images, videos, and/or messages) that they would need to passively consume to form opinions. Opinion formation is assessed across three dimensions, 1) data-type(s), 2) context, 3) and source. This work provides a theoretical basis for estimating the amount of data needed to influence a population through social media information.*

## 1. Introduction

Typical active social media usage consists of individuals posting social media information in the form of image, video, and/or message data-types to publicize their personal beliefs or thoughts. In contrast, social media viewing, or passive social media consumption, has been shown to shape an individual's perspective (i.e., opinion) [1]. To this end, individuals seek out social media information to help themselves form beliefs, opinions, or an understanding about topics [2]. These passive social media consumers are estimated to make up the large majority of online communities [3], and information seeking behaviors have been linked to passive social media consumption [4]. The present study aims to provide quantitative population thresholds for opinion formation based on individuals' self-estimates from their hypothetical passive consumption of discrete pieces of social media data-types.

The theoretical justification behind investigating individuals' hypothetical (or estimated) opinion formation thresholds as opposed to their actual thresholds, is due to inherent issues with 1) content bias [5], 2) social influence [6], and 3) different interpretations of facts associated with the same context [7]. In an attempt to rectify these three inherent issues, the current study: 1) minimizes content bias with the complete absence of physical content, 2) addresses social influence with general categories associated with distinct social media sources (e.g., like-minded vs. different-minded posting sources) and 3) provides ambiguous but discernable context categories that minimizes differences with interpretations. Although these abstractions might ameliorate the inherent issues described here, there is a sacrifice of result relevance and applicability that comes with abstracting away details. Therefore, this work is meant to provide a 'low-resolution' estimate for ratios of, or relative population averaged opinion formation thresholds, not explicitly a threshold model. However, the results from this work can be used to provide relative predictions or ratios for the amount of content that might be needed to promote a product for example, using different combinations of data-types. Furthermore, the results are intended to provide relative influence of the measured experimental dimensions instead of exact thresholds that can be taken literally.

This study has a built-in expectation that the three different data-types will result in differences with opinion formation thresholds due to the amount of information that can be presented with respect to each data-type. For example, messages and images provide static information that might be ambiguous and require additional messages or images for disambiguation. In contrast, a video is dynamic and provides rich information that can disambiguate without the need of additional videos. Therefore, we would expect that one would need less videos to form an opinion than messages or images.

Social media has become a powerful platform for exchanging information. Recent work illustrates how social media has been used to predict movie sales [8], or estimate public opinion [9, 10]. In addition, social media has been utilized to explore brand marketing strategies [11], identify "fake news" [12, 13], and disseminate health information [14]. Furthermore, social media information propagation has been used to optimize disaster relief [15, 16], and verify reputable news sources [17-21]. These studies show how social media is used as a platform for information exchange, which relies on a massive amount of active users and passive consumers to build population based insights that can in turn, influence the beliefs and opinions of individuals.

Given the influence social media information can have over an individual's opinion, it is important to find how effective different social media data-types are for opinion formation. In the present study, opinion formation is defined as the change from a neutral (naïve) state of mind to a concrete belief or perspective, based on the accumulation of evidence (i.e., pieces of data or an amount of a distinct social media data-type), resulting in either a perceived veracity or general acceptance of the material.

Thus, this work improves our understanding of how a population averaged threshold for adopting a perspective depends on different combinations of social media data-types (i.e., Images, Videos, and/or Messages) within various levels of controversy (i.e., contexts), and originating from distinct sources (i.e., like-minded or different-minded). However, it is important to note that the measured opinion formation thresholds are based on individuals' guesses about how many pieces of social media that they think, or would like to think are needed for them to form an opinion. This estimate may not be accurate, a validation set of experiments would need to be performed to confirm these self-estimates. Instead, these self-estimates provide a basis for the relative comparison of influence from our experimental dimensions (i.e., different combinations of data-types, contextual categories, and generalized sources).

Throughout this article the term opinion formation threshold is used to describe the quantitative self-estimate provided by the participants for the amount of discrete pieces of information they believe they would need to view before adopting a perspective (i.e., opinion formation). In other words, an opinion formation threshold is the participant's self-reported estimate for the number of distinct data-type(s) (i.e., Images, Videos, and/or Messages) they would need to view, in order for them to form an opinion given a context and a posting source.

Thus, the goals of the current research were to 1) identify population opinion formation thresholds for different data-types (i.e., Images, Videos, and Messages), 2) understand the influence context has over the opinion formation thresholds, 3) determine how distinct sources modify opinion formation thresholds, and 4) show how different combinations of data-types (e.g., Videos versus Videos and Messages) modulate opinion formation thresholds.

In subsequent sections, we first provide the procedure for participant acquisition, a description of the experiment, and the data analysis technique in the Methods section. Next, the Results section shows the findings from analysis and connects them to the goals of the study. Finally, the implications of this work and the future directions are discussed.

## 2. Methods

Recent evidence showing the reliability of Amazon Mechanical Turk (MTurk) data [22, 23] provides justification for the use of this platform to collect data in the present study. This previous work on the reliability of MTurk data enables the present study to expect that the use of crowdsourcing through MTurk will also provide reliable responses for population estimates, however, the results gleaned through this approach can only provide an approximation for the individual via the sample means and variances per condition. Specifically, this approach estimates opinion formation thresholds from population means, but it is important to account for the variance in a sample to generalize opinion formation thresholds.

For our study, a computerized task asked participants to enter a number associated with their estimate for discrete social media data-types (i.e., Images, Videos, or Messages) they expected to view in a static timeframe (one day) before formulating an opinion. Participants were provided with an example of a hypothetical context (i.e., None, Low, Medium, or High) corresponding to a level of controversy to frame their self-reported estimates. To avoid content bias, the participants were initially given an example of the context on the instruction page (before the start of the experiment) and were only provided with a context cue (e.g., Low) to indicate the level of controversy associated with the condition.

Participants were randomly assigned to one of 28 conditions. Participant assignment by condition is shown in Table 1. A condition consisted of a data-type combination and a context. The seven data-type combinations were: 1) Images, 2) Videos, 3) Messages, 4) Images and Videos, 5) Images and Messages, 6) Videos and Messages, or 7) Images, Videos, and Messages. The four contexts were: 1) None – no indication of a context and no cue was provided, 2) Low – a low level of controversy was inferred with a 'Low' cue throughout the experiment, 3) Medium – a medium level of controversy was inferred with a 'Medium' cue, or 4) High – a high level of controversy was inferred with a 'High' cue.

**Table 1. Participants condition assignment.**

| Data-Type Combinations | Contexts | | | |
|---|---|---|---|---|
| | None | Low | Medium | High |
| Images | 86 | 86 | 83 | 77 |
| Videos | 74 | 72 | 84 | 80 |
| Messages | 76 | 81 | 79 | 75 |
| Images & Videos | 81 | 73 | 77 | 82 |
| Images & Messages | 78 | 78 | 77 | 72 |
| Videos & Messages | 77 | 84 | 74 | 83 |
| Images, Videos, & Messages | 81 | 86 | 80 | 86 |
| **TOTALS** | **553** | **560** | **554** | **555** |

After participants were assigned a data-type combination and context (i.e., condition), they were asked to provide a response to the number of pieces of each data-type (if more than one) needed for them to form an opinion based on three source types: 1) Unspecified – source was not specified, 2) Like – the sources were like-minded, and 3) Different – the sources were different-minded.

## 2.1. Experimental population

Participants voluntarily joined the study via MTurk and were compensated with one quarter (25₵) upon completion of the study. Approximately five minutes were required to complete the study. Participants were not eligible if they were under 18 years of age, not a current resident of the United States, participated in the pilot version of this study [24] or did not regularly engage with social media.

After removal of participants with incomplete data and outlier processing, 2222 participants were included in the present analysis. In a pilot version of this study, the outlier technique was not utilized and the results did not produce meaningful conclusions [24]. The outlier technique described here is a modified version of the median absolute deviation (MAD) technique [25]. This modified MAD technique uses participants' demographic responses to Frequency ("How often do you use Social Media?") and Duration ("How much time do you spend on Social Media daily?") questions related to social media usage. Specifically, in this application the MAD technique was used to identify each participants' outlier response boundary per provided response (i.e., per sample). The two social media usage questions were re-coded into categorical variables as shown below in Table 2. It is important to note that outlier responses do not reflect a typical statistical outlier, these responses were interpreted as individuals indicating that social media information would not result in the formation of an opinion. In other words, these specific individuals do not form opinions from social media information.

**Table 2. Demographic variables.**

| Frequency | Duration |
|---|---|
| 1 = 'Once in a while' | 1 = '0-30 mins' |
| 2 = 'Once daily' | 2 = '31-59 mins' |
| 3 = 'Multiple times daily' | 3 = '1-2 hours' |
| | 4 = '2+ hours' |

The product of the two demographic variables (Frequency and Duration) was taken to provide a score (with a maximum value of 12) for each participant response (i.e., dependent on the sample). The scores were multiplied by the median of the population sample responses (for a given data-type, context, and source), to provide the participants with their individualized outlier boundary (outlier boundary = score * sample median).

If a participant's response was greater than their outlier boundary (i.e., their score multiplied by the median of the sample in question), the data point was considered an outlier and omitted from analysis. Additionally, participants' responses of '0' (zero) were excluded from analysis. These zero responses were grouped with outliers because, as described above, in this experimental paradigm, a zero response was interpreted as the individual would not form an opinion from social media data alone, and it is illogical for participants to form opinions without consuming a minimum of one piece of information. The number of data points collected for each condition across the three sources, the number of outliers, and the percentage of data removed was tallied, but is omitted here for brevity.

Prior to outlier removal, roughly the same number of participants were randomly assigned to each condition (minimal discrepancies due to MTurk parallel data acquisition). Utilizing the outlier detection technique described above, an average across all dimensions resulted in approximately 15% of the total data being identified as outliers. Outlier data was not included in the following analysis. It should be noted that the samples were distinct from conditions, each participant provided responses for three sources per condition (Unspecified, Like, and Different), resulting in three different samples per participant per condition.

## 2.2. Procedure

First, participants answered a question that screened them for their social media usage. Next, participants completed a short demographics questionnaire prior to providing their estimates for opinion formation based on data-type(s), context, and source. At the conclusion of the experiment, participants were thanked for their participation and paid for completing the study.

A short description of the different social media dimensions as shown below for data-types (Table 3), contexts (Table 4), and sources (Table 5) was provided to participants upon the instruction page of the experiment.

Participants were provided with the exact descriptions of the three distinct data-types (Table 3) utilized in this study to identify opinion formation thresholds from hypothetical social media data-types.

**Table 3. Data-type descriptions.**

| Images | still pictures, images, and drawings. |
|---|---|
| Videos | any moving pictures, animations, and videos. |
| Messages | text, a tweet, or a post on Facebook. |

A description of the assigned context (Table 4) associated with the condition was shown on the instruction page. An example of the controversy level was also provided to help guide participants towards an understanding of the scope of the contexts, while aiming to minimally bias their opinion formation thresholds.

**Table 4. Context descriptions.**

| None | no reference to controversy |
|---|---|
| Low | minimal controversy - some people form opinions. |
| Medium | controversial - many people form opinions. |
| High | highly controversial - most people form opinions. |

The source was captured through question wording (Table 5). Participants were asked the same question three times, one for each of the sources investigated.

**Table 5. Source questions.**

| Unspecified | Before you FORM an OPINION how many data types listed below would you expect to view in a day? |
|---|---|
| Like | Before you FORM an OPINION how many data types listed below would you expect to view in a day, given that the data types were *posted by people who think like you*? |
| Different | Before you FORM an OPINION how many data types listed below would you expect to view in a day, given that the data type(s) were *posted by people with different viewpoints*? |

Images were defined as data-types that include still pictures, images, and drawings. Videos were defined as data-types that include any moving pictures, animations, or sequence of images. Messages were defined as data-types that include only text (e.g., micro-texts or posts). Context labeled None, indicated no reference to controversy; context labeled Low was associated with minimal controversy (e.g., some people would form an opinion about the information); Medium was associated with some controversy (e.g., many people would form an opinion about the information); and High was associated with much controversy (e.g., most or all people would form an opinion about the information). Each participant was asked to estimate their opinion formation from all three sources. The three sources were either 1) unspecified to the participant labeled Unspecified, or 2) from like-minded individuals labeled Like (e.g., posted by those who have similar viewpoints to the participant; in-group), or 3) from different-minded individuals labeled Different (posted by those who have different viewpoints from the participant; out-group).

## 3. Results

The analysis procedure involved, 1) testing each sample associated with a single context, data-type presentation, and source for normality (i.e., if the sample comes from a normal distribution with an unspecified mean and standard deviation), 2) conducting Quantile-Quantile plot tests per sample to assess the most likely underlying distribution (determined to be log-normally distributed), 3) performing a log-transform on the log-normally distributed data samples (to allow for standard parametric testing with population statistics), and 4) perform mixed-measures analysis of variance (ANOVA) to ascertain differences between sample population statistics. This analytical approach allowed for an understanding of relevant differences between samples while adhering to the assumptions of parametric tests.

### 3.1. Data transformation and analyses

Throughout this subsection, the data refer to the conglomeration of all data samples across the different conditions. Appropriate testing was conducted on each individual sample, and not the dataset as a whole.

Jarque-Bera (JB) goodness-of-fit tests were initially used to determine if the data came from unspecified normal distributions. The JB test results indicated that the data was not normally distributed. However, the data fit a log-normal distribution, confirmed with exhaustive Quantile-Quantile plot (Q-Q plot) testing. Therefore, a power transform (natural log) resulted in normally distributed data; confirmed with additional post-transform JB tests. The following parametric analyses were performed on the log transformed data with the final opinion formation thresholds (i.e., means) reported as the inverse log transform of the statistics taken in the log transformed space (i.e., statistics were transformed back into the original non-transformed space).

Separate mixed-measures analysis of variance (ANOVA) were performed for each of the social media data-types (i.e., Images, Videos, and Messages) across the different combination types (i.e., Single vs. Multimedia conditions) as the between-participants factor, context (i.e., None, Low, Medium, and High) as a between-participants measure, and social media source type (i.e., Unspecified, Like, and Different) as the within-participants factor. In some instances, assumptions of sphericity were violated and a Huynh-Feldt epsilon statistic is reported where appropriate.

### 3.2. Images data-type

There was a significant interaction between source type and combination type, $F(6, 1912) = 4.94$, $p < .001$, $\eta p2 = .015$, $\varepsilon = .96$. This relationship indicates that more images were needed to form opinions when the source was unspecified and participants were asked to estimate multiple social media data-types together (compare Figure 1A to 1B, 1C, and 1D). There were no additional significant interactions to report for these analyses.

There was a significant main effect of source on the approximate number of images required to form an opinion from the Images data-type, $F(2, 1912) = 151.14$, $p < .001$, $\eta p2 = .14$, $\varepsilon = .96$. Bonferroni-corrected pairwise comparisons indicated a significant difference between source types. The data show that an unspecified source resulted in significantly greater opinion formation thresholds across all contexts when compared to other sources (Unspecified vs. Like and Unspecified vs. Different), within the same combination type (compare within Figures 1B, 1C, or 1D Unspecified to Like and Different), and the single media type for the Images data-type (compare Unspecified between Figures: 1A to 1B, 1A to 1C, and 1A to 1D). In addition, an unspecified source resulted in a significantly greater amount of images needed to form an opinion when compared to like-minded sources (Unspecified vs. Like, $p < .001$), and different-minded sources (Unspecified vs. Different, $p < .001$).

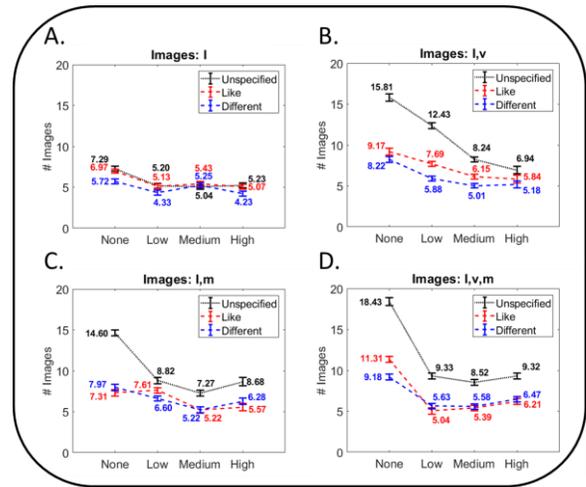

**Figure 1. Images: population mean values per context, source, and combination type. Sample means are color-coded values with error bars indicating standard error of the mean (SEM). Black shows no indication of sources (Unspecified), red like-minded sources (Like), and blue different-minded sources (Different). The y-axes show the number of images for population estimates. The x-axes show contexts. 1A. Images alone (single media type). 1B. Images paired with videos – I,v (multimedia). 1C. Images and messages – I,m (multimedia). 1D. Images, videos, and messages - I,v,m (multimedia).**

There was a significant main effect of context on the approximate number of images required to form an opinion from social media data, $F(3, 956) = 11.88$,

$p < .001$, Ƞp2 = .04. Bonferroni-corrected pairwise comparisons indicated significant differences between the number of images to form an opinion when the context was not indicated and a low level of controversy (None vs. Low, $p < .001$). A similar result was found for medium and high levels of controversy (None vs. Medium, $p < .001$; None vs. High, $p < .001$). Furthermore, an unspecified context resulted in significantly more images to form an opinion across multimedia combination types (Figures 1B – 1D).

There was a significant main effect for the combination of different data-types on the number of images required to form an opinion, $F(3, 956) = 10.02$, $p < .001$, Ƞp2 = .03. Bonferroni-corrected pairwise comparisons indicated a significant difference between the number of images required to form an opinion for a single media type (Images alone) and multimedia types (Images in combination with the other data-types). Compare Figures 1A to 1B: [I vs. I,v], $p < .001$; 1A to 1C: [I vs. I,m], $p = .002$; and 1A to 1D: [I vs. I,v,m], $p < .001$). All multimedia combination types resulted in a significant increase in the participants' responses for the amount of images to form an opinion, over the single media type.

### 3.3. Videos data-type

There was a significant three-way interaction between source type, combination type, and context for the Videos data-type, $F(18, 1820) = 2.42$, $p = .001$, Ƞp2 = .02, ε = .92. A significant interaction was also identified between source type and combination type for the population averaged number of videos needed to form an opinion, $F(6, 1820) = 3.65$, $p = .002$, Ƞp2 = .01, ε = .92. This relationship indicates that more videos were needed to form opinions when the source and context were not indicated (Figure 2).

Similar to Images data-type results, a significant main effect of source on the approximate number of videos needed to form an opinion, $F(2, 1820) = 130.00$, $p < .001$, Ƞp2 = .13, ε = .92. Bonferroni-corrected pairwise comparisons indicated that significantly more videos were required to form an opinion when comparing unspecified sources to like-minded sources population averages (Unspecified vs. Like, $p < .001$), and to different-minded sources population average (Unspecified vs. Different, $p < .001$). The significantly greater population estimate of opinion formation threshold from an unspecified source is shown in Figure 2.

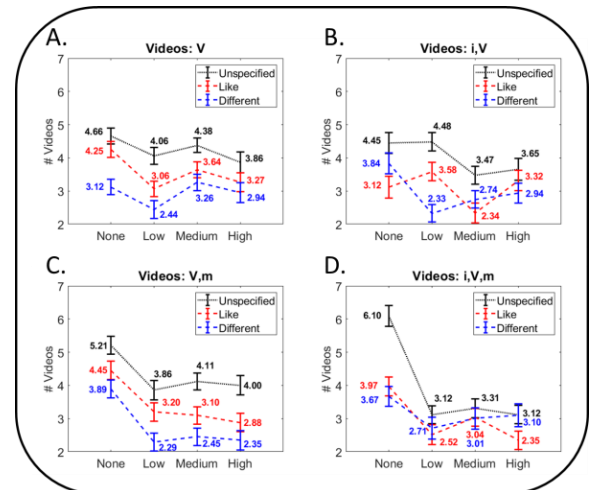

**Figure 2. Videos: population mean values per context, source, and combination type. Sample means are color-coded values with error bars indicating standard error of the mean (SEM). Black shows no indication of sources (Unspecified), red like-minded sources (Like), and blue different-minded sources (Different). The y-axes show the number of videos for population estimates. The x-axes show contexts. 1A. Videos alone (single media type). 1B. Videos paired with images – i,V (multimedia). 1C. Videos and messages – V,m (multimedia). 1D. Videos, images, and messages - i,V,m (multimedia).**

There was a significant main effect of context on the population averaged number of videos reported to form an opinion for the Videos data-type, $F(3, 910) = 9.12$, $p < .001$, Ƞp2 = .01. Bonferroni-corrected pairwise comparisons indicated that significantly more videos were required to form opinions when the controversy surrounding the information was not known, compared to the low controversy level (None vs. Low, $p < .001$), the medium controversy level (None vs. Medium, $p = .003$), and the high controversy level (None vs. High, $p < .001$).

There were no main effects of combination type to report for these analyses. Unlike the Images data-type, there was not a significant difference between single and multimedia combination types for the Videos data-type (compare Figures 2A to 2B, 2C, and 2D).

### 3.4. Messages data-type

There was a significant interaction between source and combination type, $F(6, 1886) = 3.77$, $p = .001$, Ƞp2 = .01, ε = .95. This relationship indicated that more messages were needed by the population on average to form opinions when the source was unknown and multimedia combination types were considered (Figure 3). The population averages across conditions clearly show the significant differences between the single media type to the multimedia combination types (compare Figure 3A to 3B, 3C, and 3D). There were no additional significant interactions to report for these analyses.

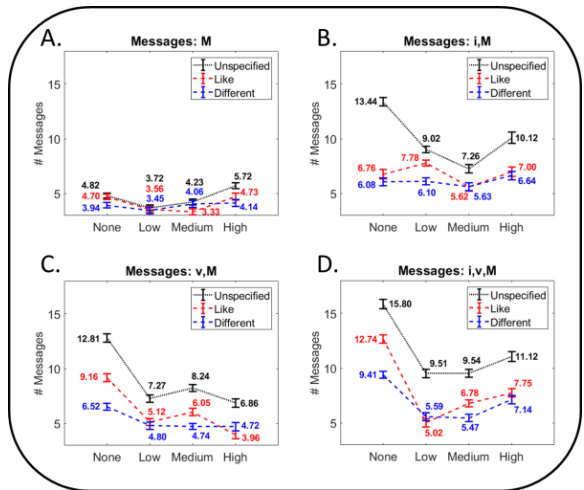

**Figure 3. Messages: population mean values per context, source, and combination type. Sample means are color-coded values with error bars indicating standard error of the mean (SEM). Black shows no indication of sources (Unspecified), red like-minded sources (Like), and blue different-minded sources (Different). The y-axes show the number of messages for population estimates. The x-axes show contexts. 1A. Messages alone (single media type). 1B. Messages paired with images – i,M (multimedia). 1C. Messages and videos – v,M (multimedia). 1D. Messages, images, and videos - i,v,M (multimedia).**

There was a significant main effect of source on the population averaged number of messages reported to form an opinion, $F(2, 1886) = 138.59$, $p < .001$, $\eta p2 = .13$, $\varepsilon = .95$. Homologous to the Images and Videos data-types, Bonferroni-corrected pairwise comparisons indicated that significantly more messages were required to form opinions with unspecified sources compared to like-minded (Unspecified vs. Like, $p < .001$), and different-minded sources (Unspecified vs. Different, $p < .001$). The results show that there was not significant differences between sources for the single media type (Figure 3A). Similarly, there was not a significant difference between like-minded (Like) and different-minded (Different) sources for multimedia combination types (Figures 3B – 3D).

There was a significant main effect of context on the population averaged number of messages reported to form an opinion, $F(3, 943) = 28.47$, $p < .001$, $\eta p2 = .08$. Comparable to the results from the Images and Videos data-types, Bonferroni-corrected pairwise comparisons for the Messages data-type indicated that significantly more messages were required to form an opinion when the controversy surrounding the information was not indicated, compared to the low controversy level (None vs. Low, $p < .001$), and the medium controversy level (None vs. Medium, $p = .003$), but not the high controversy level (None vs. High, $p = .490$).

There was a significant main effect of presentation type on the population averaged number of messages reported to form an opinion for the Messages data-type, $F(3, 943) = 5.88$, $p < .001$, $\eta p2 = .02$. Bonferroni-corrected pairwise comparisons indicated a significant difference between a single media type (M = 1.40, SE = .05) and the three multimedia combination types ([M vs. i,M], $p < .001$; [M vs. v,M], $p < .001$; [M vs. i,v,M, $p < .001$). These results suggest that the multimedia combination types result in a significant increase in messages to form opinions over the single media type (compare Figure 3A to 3B – 3D).

### 3.5. Opinion formation thresholds

These results indicate how data-type, source, context, and combination type influence the population averaged number of images, videos, or messages respectively, which are needed to form opinions strictly from social media data. Given that the samples from each condition per data-type have variance, an intuitive conclusion from the results is to estimate the opinion formation thresholds to be the means of the samples (see values in Figures 1 – 3). Furthermore, the results suggest that multimedia combinations of Images and Messages data-types elicit a significantly greater amount of respective data-type to form opinions over the single types (i.e., Images alone or Messages alone). In contrast, the multimedia combinations of the Videos data-type demonstrate no such effect. Therefore, the results show that multimedia combinations do not have an impact on the Videos data-type, but these combinations yield significant increases in opinion formation thresholds for the Images and Messages data-types.

## 4. Discussion

The present study addresses how factors influence opinion formation through a crowdsourced experiment, which collects participants self-estimated opinion formation thresholds across different dimensions of social media information (i.e., data-type, context, and source). The goals of the current study were to 1) identify opinion formation thresholds associated with passive social media data consumption (passive viewing) across the dimensions of data-type (i.e., Images, Videos, and Messages), source (i.e., Unspecified, Like, and Different), combination type (single vs. multimedia), and context (i.e., None, Low, Medium, and High), 2) examine opinion formation threshold differences between single and multimedia combination types (e.g., Images vs. Images & Videos), 3) determine how context, represented as different levels of controversy influence opinion formation thresholds, and 4) understand how opinion formation thresholds adjust with source type.

The results from this work provide a set of population estimates that are valid in comparison between samples for the amount of social media data needed to impact the opinions of individuals. The relevant findings from this work are: 1) population averaged opinion thresholds identified through exhaustive statistical analysis that represent a population averaged self-guess of true thresholds, 2) influence from unspecified sources significantly increased the estimated threshold relative to other sources independent of data-type, 3) influence from the abstracted contexts significantly depended on data-type combination, and 4) opinion formation threshold estimates were significantly greater when comparing multimedia to single media for Images and Messages, but not for Videos.

Across all experimental dimensions, it was clear that when the social media sharing source was not specified to the participant (Unspecified), the threshold estimates were significantly higher compared to like-minded (Like) and different-minded (Different) sources. This is perhaps due to the inherent uncertainty associated with unspecified sources, which can yield lower trust in the information, resulting in the individual's reported need to view more of a select data-type before enabling individuals to form opinions. Conversely, evidence has shown that an individual may conform their opinions to their like-minded or different-minded peers with significantly less data [26]. For example, if a like-minded peer of an individual is quick to form an opinion based on information presented from videos shared on social media, that individual may also be quick to form an opinion due to social cohesion with that peer. In an attempt to avoid this social cohesion bias, we took measures by abstracting the source of any personal ties and resolving it to just a general category of like-minded, different-mind, and unspecified. However, it is possible that participants projected their own perceptions of the source categories onto their social networks, so further testing would need to be conducted to resolve this matter.

Similar to the generalized categories of source, specific content informed by the contexts was intentionally abstracted away to minimize a bias of previous experiences associated with actual images, messages, or videos. Historically, it has been shown that content will affect an individual's opinions based on their personal experiences [26]. Due to the complexities that connect content to personal experiences, the present study did not utilize content. To further understand how context modifies opinion formation, content should be cautiously introduced.

Analysis of context revealed that when controversy was absent (None), thresholds across most dimensions were significantly greater than the Low, Medium, and High controversy cases. Intuitively, this appears to suggest that when an individual has less information surrounding the social media data (context was unspecified), a significantly greater amount of data is needed before an opinion can be formed. However, our results suggest that the specified contexts (i.e., Low, Medium, and High) did not have an intuitive monotonically increasing relationship, from Low to High. Here is an example to demonstrate this intuitive pattern; if the price of a product is related to our specified contexts, one might expect that fewer pieces of data would be needed by an individual to make a purchase decision for a low cost item. Whereas, the same individual might need a greater amount of data to make the purchase decision if the item in question met their subjective criteria for a medium cost item, and yet even more data for a high cost item. In contrast, our results did not exhibit this pattern except in select cases (Messages: IVM – Unspecified and Like), but the differences did not reach significance. Therefore, we suspect that our categorical contexts were too general to capture this intuitive fine resolution trend. Future work will explore more specific contextual scenarios (e.g., cost of items or social events) that should yield the intuitive pattern mentioned above.

The results show a significant increase in opinion formation threshold for multimedia relative to single media for Images and Messages data-types, but not for Videos. This could imply that when individuals are given multiple different data-types together, an interaction between diversity of data-type and their internal information processing drives up the number of images and messages required to form an opinion. However, the number of videos required to form an opinion did not show dependence on single or multimedia combinations, possibly due to the richness of information available in a videos relative to images or messages.

In summary, opinion formation thresholds for the Images and Messages data-types were similar, however, thresholds for the Videos was consistently less across the dimensions. This phenomenon can be explained by qualitatively approximating the amount of information inherently associated with each of the data-types. At the very least, the Videos data-type consists of a sequence of images, which can easily translate to more information available than the Images data-type. In contrast, Images and Messages data-types might provide approximately the same amount of information, resulting in similar relative thresholds across the dimensions. Given that the trends discovered through rigorous statistical analysis support this qualitative approximation, this appears to be a reasonable conclusion for the patterns identified, and supporting results were found in literature [2].

These results regarding population opinion formation thresholds in the presence of single and multimedia data-types can be of immense importance in many areas of sociology and complex networks. In fact, results of this type can feed directly into stochastic models that simulate opinion spread through society. Examples include dosage based models of opinion spread [27]. Furthermore, there exist computational models that deal with individuals that are particularly stubborn and difficult to change [28-30], similar to the noted population of outliers that would not form an opinion. Finally, future models can be developed using the information gained here; the results showing different thresholds for different data-types, sources, and contexts could be used to build new variants of previously studied models to capture specific facets of social interactions.

Future work will investigate the relationship between opinion formation, specific content, and contextual scenarios to build a more complete understanding of how social media data can shape opinions of individuals.

## Acknowledgements

This research was sponsored by the United States Army Research Laboratory (USARL) under Cooperative Agreement Numbers W911NF-17-2-0088 (Oak Ridge Institute for Science and Education - ORISE) and W911NF-09-2-0053 (Network Science Collaborative Technology Alliance - NSCTA). This research was supported in part by Rensselaer Polytechnic Institute as an appointment to the Student Research Participation Program at USARL administered by ORISE through an interagency agreement between U.S. Department of Energy and USARL, and the Research Associateship Program at USARL administrated by ORAU.

## Disclosure Statement